# Interpretable machine learning in Physics

**Christophe Grojean**[1,2], **Ayan Paul**[1,2,*], **Zhuoni Qian**[3], **and Inga Strümke**[4]

[1]Deutsches Elektronen-Synchrotron DESY, Notkestr. 85, 22607 Hamburg, Germany
[2]Institut für Physik, Humboldt Universität zu Berlin, 12489 Berlin, Germany
[3]School of Physics, Hangzhou Normal University, Hangzhou, Zhejiang, 311121, China
[4]Department of Computer Science, Sem Sælandsvei 9, Norwegian University of Science and Technology, Norway
*e-mail: christophe.grojean@desy.de, ayan.paul@desy.de, zhuoniqian@hznu.edu.cn, inga.strumke@ntnu.no

Adding interpretability to multivariate methods creates a powerful synergy for exploring complex physical systems with higher order correlations while bringing about a degree of clarity in the underlying dynamics of the system.

To make the intricacies of machine learning models intelligible, explainability of these models are becoming increasingly important from the ethical, legal, security, transparency and robustness standpoints, especially when they are used to make decisions that can affect human lives [1]. This has fueled the development of interpretable machine learning [2] or explainable AI [3] methods that try to eradicate the problems of a *black-box* approach to multivariate analyses. Needless to say, novel opportunities for the use of these algorithms to study the structure and dynamics of physical systems have been ushered in too. However, the dawn of explainable AI in physics has been mostly relegated to explaining how neural networks analyze experimental data rather than a move towards more interpretable machine learning frameworks that facilitate a better understanding of the dynamics learnt by the models and build confidence in the ensuing analyses [4].

For instance, particle collisions at the Large Hadron Collider have rich kinematic structures that allow for study of the underlying dynamics governed by the forces of Nature. This has facilitated, amongst others, the discovery of the Higgs boson and matter-antimatter asymmetry in which machine learning algorithms have played a central role [5]. While the outcomes of these analyses have been well understood, the inner workings of the machine learning models have remained obscure because of the inexplicability of the parameter space or topology of the models. It is this black-box approach to statistical analyses has kept at bay the applicability of machine learning in a wider scope of physics since interpretability is of prime importance, i.e., where not only the results but also a well-rounded interpretation of the physical hypotheses needs to be understood.

In an attempt to move towards more interpretable machine learning frameworks, we present a beginner's guide to building multivariate analyses that can be made interpretable. The ability to comprehend the hierarchy of variable importance in performing a classification or regression creates a paradigm shift in the understanding of the dynamics that govern physical processes, bringing about a degree of clarity and confidence in the comprehension of the dynamics learnt by the machine learning models.

## Decision Trees and Neural Networks

By far the two most common machine learning frameworks used for modeling non-linear relationships in scientific data are decision trees and neural networks. As a multivariate method decision trees, while being easily interpretable, are not as capable as neural networks. Being universal function generators neural networks, generalize much better given noisy data but at the cost of being black boxes since their parameters do not provide any insight into how the input variables are connected with the output variables. More complex frameworks made of ensembles of decision trees, like boosted decision trees (BDTs) [6] and random forests [7], successfully address the limitations of decision trees but some of the interpretability is lost in boosting and much of it is lost for random forests. However, they retain a distinct advantage over neural networks because they are typically faster to train and perform equally well or better for tabular data, given that it is not too noisy. Of course, there are cases where the use of neural networks is necessary, e.g., image recognition, sequence to sequence analyses, etc.

## Model Interpretations

An increase in the complexity of models necessary to deal with more complex datasets correlates to a decrease in the interpretability of the models. So as to reintroduce an element of interpretability of trained models, post hoc methods have been proposed that examine the machine learning algorithms and try to find reasons why a decision is being made. Broadly classified, they either fall into the category of *local* methods that explain every outcome in terms of the input variables, or into the category of *global* methods that interpret the model as a whole and determine the overall "variable importance" which builds a hierarchy of the importance of the input variables in determining the output. Some tools for interpretability are mathematically less complex and computationally less demanding. LIME [8] is a *local* method that creates surrogate interpretable models locally to explain each prediction. For *global* interpretations through analysis of variable importance, permutation or Gini Index based methods can be used.

As a more mathematically robust method, the Shapley decomposition is a solution concept for transferable utility games in Coalition Game Theory that fairly distributes the payoff of a game among its players and is provably the only solution satisfying the four favourable axioms of efficiency, additivity, symmetry and null player. In a multivariate analysis, the input variables can be considered as the players and the regression or classification performed by the machine learning algorithm can be considered as the payoff allowing for the inference of relative importance of each input variable in determining the output variable(s). This inference can be made *locally* for each prediction, and a *global* hierarchy of variable importance can be constructed as well for the system under study [9] to explain ensembles of trees or neural networks. SHapley Additive exPlanations (SHAP) is an popular tool sharing some basic principles with LIME that interprets machine learning models using Shapley values.

## Simple example from particle physics

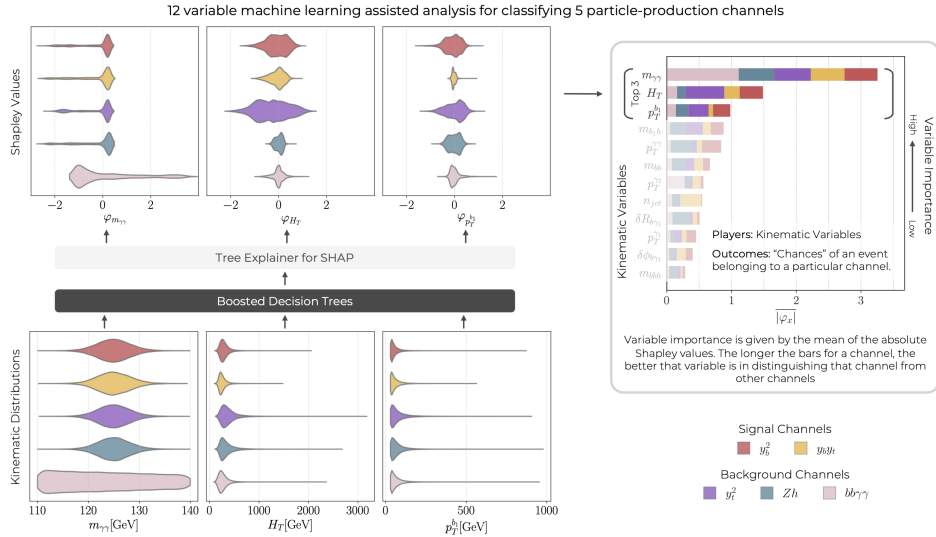

**Figure 1.** *From physics to interpretations:* a schematic diagram of the flow from kinematic distributions to **local** interpretations given by Shapley value distributions, $\varphi_x$, using a Boosted Decision Tree and SHAP. This leads to the hierarchy of variable importance, represented by $\overline{|\varphi_x|}$, of the kinematic variables facilitating a **global** interpretation of the BDT and providing insights into the dynamics driving the different channels

An example of an interpretable analysis can be constructed for the measurement of the Yukawa coupling of the Higgs to the bottom quarks through the production of the Higgs in association with a bottom quark pair [10]. This measurement faces the challenge of extracting an extremely small signal from kinematically similar backgrounds which cannot be separated using traditional methods that are based on kinematic cuts. A machine learning model, for instance, a fitted BDT, makes the task much more feasible but at the cost of making the analysis less transparent.

Understanding the dynamics is made possible through the attribution of variable importance once one uses Shapley values to interpret the trained BDT. In Fig. 1 (bottom panels) we show the predicted kinematic distributions for the signal and background processes which are mostly similar. This does not prove to be a hurdle for a BDT since it can classify them very effectively leveraging *higher-order correlations*. The deformation of the kinematic distributions into the Shapley value distributions (top panels) incorporates these correlations learnt by the machine learning model, hence, providing *local* interpretability of how the channels are separated. Finally, the plot in the right panel with the mean of the absolute Shapley values that quantifies how well the machine learning model separates the different channels overall leveraging a particular variable, lays out the hierarchy of variable importance and how the variables "play a cooperative game" to orchestrate the outcome providing a *global* interpretation of the model. Shapley values are rapidly gaining favour as a tool for interpreting machine learning models used in particle physics analyses [11–13].

## Future of Interpretability

When a black-box approach to multivariate analyses is taken, the predictive power of a machine learning model does not guarantee a concordance between the dynamics driving a physical process and the mathematical structure statistically modelled by the learning algorithm. Amongst the plethora of possible configurations the parametric solutions of the machine learning algorithm might assume, several solutions can easily be unphysical or simply be correct in only a subset of the states that the actual physical system can explore. Hence, the knowledge of how and why a machine learning model predicts a certain outcome for a physical system provides confidence that the results are not, by fluke, only *seemingly* correct, but rather stand on robust grounds of well understood physical theories, like the Standard Model of particle physics.

The combination of machine learning with post hoc model interpretations has found applicability in various fields like complex systems, climate science, space science, astrophysics, astrobiology, etc. Wider acceptance and applications in other branches of physics is just a matter of time as the combination brings to our disposal the power to probe complex multivariate and non-linear systems while not having to forsake the ability to understand the path to the solutions that we propose.

**Acknowledgements**
This work benefited from support by the Deutsche Forschungsgemeinschaft under Germany's Excellence Strategy EXC 2121 "Quantum Universe" – 390833306. The work of A.P. is funded by Volkswagen Foundation within the initiative "Corona Crisis and Beyond – Perspectives for Science, Scholarship and Society". I.S. is grateful to the Norwegian Research Council for support through the EXAIGON project – *Explainable AI systems for gradual industry adoption* (grant no.304843).

**Author Contributions**
The authors contributed equally to all aspects of this article.

**Competing Interests**
The authors declare no competing interests.